\documentclass[english,aps,prstper,reprint,showpacs,titlepage,longbibliography]{revtex4-2}   

\usepackage[T1]{fontenc}	
\usepackage[latin9]{inputenc}	
\usepackage{geometry}		
\usepackage{graphicx}
\usepackage[above,below]{placeins}	
\usepackage{times}
\usepackage{tabularx}

\usepackage{hyperref}  
\hypersetup{colorlinks=true,urlcolor=blue,citecolor=blue,linkcolor=blue}   
\usepackage{enumitem}          
\setlist{nosep}                 

\newenvironment{iquote}
    {\vspace{-.5\baselineskip}\itshape\list{}{\leftmargin=0.2in\rightmargin=0.2in}%
    \item\relax}
    {\endlist\vspace{-.5\baselineskip}}

\newcommand{\myquote}[1]{\vspace{-.1\baselineskip}\begin{iquote} 
#1\end{iquote}\vspace{-.1\baselineskip}}

\begin{document}

\begin{titlepage}

\title{Representational differences in how students compare measurements}

\author{Gayle Geschwind$^{1,2}$}
\author{Michael Vignal$^{1,2}$}

\author{H. J. Lewandowski$^{1,2}$}

\affiliation{$^{1}$Department of Physics, University of Colorado, Boulder, Colorado 80309, USA}
\affiliation{$^{2}$JILA, National Institute of Standards and Technology and University of Colorado, Boulder, Colorado 80309, USA}


  \begin{abstract}

  Measurement uncertainty plays a critical role in the process of experimental physics. It is useful to be able to assess student proficiency around the topic to iteratively improve instruction and student learning.  For the topic of measurement uncertainty, we developed an assessment tool called the Survey of Physics Reasoning on Uncertainty Concepts in Experiments (SPRUCE), which aims to assess students' knowledge, and use of, a variety of concepts related to measurement uncertainty. This assessment includes two isomorphic questions focused on comparing two measurements with uncertainty. One is presented numerically and the other pictorially. Despite the questions probing identical concepts, students answer them in different ways, indicating that they rely on distinct modes of representation to make sense of measurement uncertainty and comparisons.  Specifically, students score much higher on the pictorially represented item, which suggests possible instructional changes to leverage students' use of representations while working with concepts of measurement uncertainty. 
  
 \clearpage
  \end{abstract}

  \maketitle
\end{titlepage}

\section{Introduction \& Background \label{sec:intro} }
\vspace{-4mm}

All measured quantities have associated uncertainties, making measurement uncertainty a crucial aspect of experimental physics. Using measurement uncertainty correctly is essential for interpreting measurements, presenting results, and drawing reliable conclusions based on those results. The Effective Practices for Physics Programs (EP3) Guide \cite{EP3} also emphasizes the significance of learning measurement uncertainty techniques as taught in physics laboratories. Despite its critical role, students frequently struggle with concepts and practices surrounding measurement uncertainty, including propagation of error, comparison of measurements, calculating standard deviations and standard errors, and taking several measurements to get a distribution of results, even after taking a course emphasizing these areas \cite{Campbell2005, Holmes2015, Holmes2015_3, Stein2018, Quinn2018, Kok2022}.

As part of efforts to improve student learning of measurement uncertainty, we have developed a new research-based assessment instrument (RBAI) called the Survey of Physics Reasoning on Uncertainty Concepts in Experiments (SPRUCE) \cite{Pollard2021, Vignal2023}.  SPRUCE is an online assessment intended to be utilized in a pre-post format allowing instructors to measure the impact of a course on students' proficiency with concepts and practices of measurement uncertainty. We developed SPRUCE using the framework of Evidence-Centered Design (ECD) \cite{Mislevy2005}, a robust method of creating and validating an RBAI. Although validation is an ongoing project (with a future paper in progress), SPRUCE still offers a wide variety of insights into how students handle measurement uncertainty. Its design provides instructors with their students' progress along 14 dimensions referred to as Assessment Objectives (AOs) \cite{Vignal2022} after one term of a laboratory class.

AOs are \textit{``concise, specific articulations of measurable desired student performances regarding concepts and/or practices targeted by the assessment} \cite{Vignal2022}\textit{.''}  AOs are similar to course learning goals and are essentially the constructs the assessment aims to measure. We developed the SPRUCE AOs with input from introductory laboratory instructors to determine which aspects of measurement uncertainty they find important and want their students to learn in their courses \cite{Pollard2021}. These AOs then aided in writing the SPRUCE assessment items: each item on SPRUCE addresses at least one of these objectives. In this way, we focused the scope of SPRUCE to topics instructors frequently deem important to their introductory laboratory courses.

Here, we examine one particular SPRUCE AO: \textit{Determine if two measurements (with uncertainty) agree with each other}. SPRUCE has two isomorphic questions for this objective. First, the assessment presents students with numerical measurements and asks about agreement between these measurements. Then, later in the assessment, with several questions in between, a similar question appears with the data represented pictorially, as symbols with error bars. Students are not explicitly informed about the relationship between these two items. This allows us to probe how students are able to compare measurements when presented with the same data with two different representations.

Existing literature has explored the use of multiple representations while students problem solve \cite{Kohl2005, Kohl2006, Kohl2007, Kohl2008, Susac2017}. For example, Kohl et al. found that students frequently view a mathematical problem and a pictorial problem as `opposites,' where students consider pictorial problems as more aligned with ``concepts,'' which are frequently treated distinctly from numerical problems. Further, they found statistically significant differences in performance based on different representations of isomorphic problems on homework and quizzes. Students tended to perform worse on problems in a mathematical or numerical format than with problems in other formats (e.g., pictorial, verbal, or graphical)\cite{Kohl2005}.       

The work presented here aims to identify whether student performance in comparing measurements similarly depends on representation. To do this, we will answer the following research questions.
\begin{itemize}

    \item  Do students respond differently to questions about comparing measurements when presented with different representations? 
    \item How do students reason about comparing measurements when presented with different representations?
\end{itemize}

\vspace{-5mm}  

\section{Methodology \label{sec:methods}}

\vspace{-4mm}  

We use a mixed methods approach, as the data collection and the analysis involve qualitative and quantitative components. To study students' handling of measurement uncertainty, we administered SPRUCE in a pre-post online format during the Fall 2022 semester in 12 courses at eight institutions (See Table \ref{tab:types}). We received 670 valid post-instruction responses after we removed responses from students who did not consent to have their data used for research, did not correctly answer the filter question, or did not answer both items of interest.

    

\begin{table}[t]
    \centering
    \caption{Institutions and student responses in the dataset after removing student answers for incorrect filter/nonconsent to research}
    \label{tab:types}
    \begin{tabularx}{\linewidth}{ccc}\hline\hline
        Number of Institutions & Institution Type & Number of post responses \\ \hline
        1 \: & 2 Year & 7\\ 
        1 \: & 4 Year & 7 \\ 
        1 \: & Master's & 39 \\ 
        5 \: & PhD & 617 \\ 
    
    \hline \hline
    \end{tabularx}
\end{table}

We also conducted interviews during the Fall 2022 semester. These interviews aimed to determine whether students interpreted all of the items on SPRUCE as intended, as well as to probe student reasoning for each answer option on the assessment. Students were recruited from seven courses at four institutions (2-year, Master's and PhD granting) already participating in the administration of SPRUCE during this semester.  Each of the 27 interviews conducted lasted approximately one hour and students were compensated for their time. Interviewers (two of the authors) observed as students completed SPRUCE and inquired about students' reasoning for each answer selected, as well as about why they did not select certain answer options. The interviews were audio/video recorded for future reference. Analysis of these interviews consisted of taking notes during interviews and transcribing student quotes as needed. 

\begin{figure}[!ht]
    \centering
    \includegraphics[width=.45\textwidth]{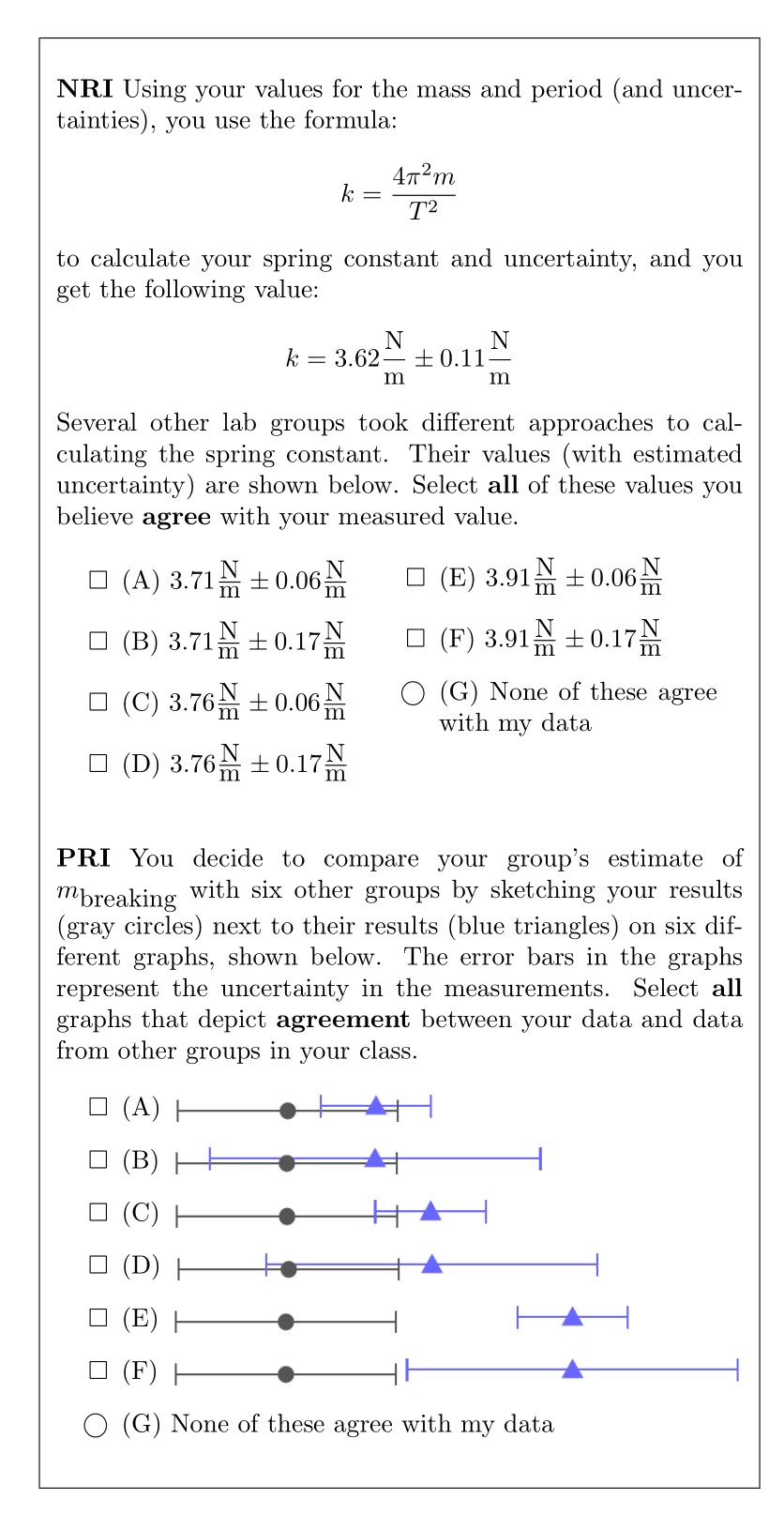}
    \caption{Two Isomorphic Items on SPRUCE. These items probe student understanding of measurement comparisons with uncertainty by presenting the same data in two different representations - a numerically represented item (NRI) and a pictorially represented item (PRI). The students first encounter the NRI and then, after answering several unrelated questions, they encounter the PRI. Note that the answer options on the PRI are in a different order when presented to students (DAEBFCG) than shown here; we present them in the same order as the answer options for the NRI in this paper for ease of understanding.}
    \label{fig:exptfigure}
\end{figure}

For the analysis, we focused on the responses to two isomorphic multiple-response items, focusing on both the difficulty \cite{Engelhardt2009, Ding2009} of these items and student reasoning for their answers to both. The first item, as shown in the upper half of Fig. \ref{fig:exptfigure}, presents students with their `own' numerical data (with uncertainty) for a measurement of a spring constant; they are then asked to select all answer choices of numerical data (means with uncertainties) that agree with their measurement. The second item presents these similar data in pictorial form, as shown in the lower half of Fig. \ref{fig:exptfigure}. For brevity, we will refer to the numerically represented item as NRI and the pictorially represented item as PRI for the remainder of this paper. Students receive credit on these multiple response items by answering with the combination `ABCD' or `ABCDF,' based on expert responses. The uncertainties in both items represent the standard error; therefore, overlap or near overlap of the error bars is required for agreement. No other answer combinations earn credit, and no partial credit is awarded for these items. Note that we changed the order of answer options for the PRI for this paper to make discussion of the items easier.

\vspace{-4mm}
\section{Results \& Discussion \label{sec:resultsanddiscussion}}

\vspace{-4mm}
\subsection{Overall difficulty scores}
\vspace{-3mm}
While laboratory instruction commonly focuses on measurement comparison \cite{Pollard2021}, low scores on both of these items at the end of the term indicate persistent student difficulties in handling comparison with uncertainties. Students score an average of (25 $\pm$ 3)\% on the NRI and an average of (40 $\pm$ 4)\% on the PRI on the post-test, with the error indicating 95\% confidence interval. These scores indicate that, while not many students answered these items correctly, students answered the PRI correctly more often. We conducted a Mann-Whitney U test (a nonparametric test for independent measures) \cite{Mann1947} to determine if this represents significant statistical difference, and found the p-value for these items as $p = 2.1 \times 10^{-8}$, indicating a statistically significant difference in student performance on these items.Additionally, we calculated the effect size to compare these two items using Cohen's $d$ \cite{Kirk2008, Cohen1988}, finding $d = 0.31 \pm 0.05$, showing a moderate effect size. 

We also calculated the Pearson coefficient to determine the correlation between the two items. The Pearson coefficient varies between $r=-1$ and $r=1$, where a more positive coefficient indicates a stronger positive correlation \cite{Engelhardt2009}. Anything above $r \approx 0.30$ indicates a fairly significant positive correlation. For these items, we find $r = 0.45 \pm 0.04$, which shows a fairly significant correlation in that if a student correctly answered one item, they are more likely to have correctly answered the other. However, the correlation is not perfect ($r=1$): many students correctly answer only one of these items. The number of students who answered each question correctly is presented in Table \ref{tab:correct}. Only about half of the students who correctly answered the PRI also correctly answered the NRI, but about 75\% of the students who correctly answered the NRI also correctly answered the PRI. This suggests that students who are able to reason through the numerically presented data seem better equipped to handle the pictorially presented data, but the reverse is not true on average.

\begin{table}[t]
    \centering
    \caption{Number of students who answered the NRI, PRI, or both correctly [N = 670]; error shown as 95\% confidence interval}
    \label{tab:correct}
    \begin{tabularx}{\linewidth}{r X X X }\hline\hline
         & Only NRI & Only PRI & Both Correct \\ \hline
        Number of Students \: & 38 &  134 & 131\\ 
        Percent of Students \: & 6 $\pm$ 1 & 20 $\pm$ 3 & 20 $\pm$ 3 \\

    \hline \hline
    \end{tabularx}
\end{table}

We turn to the qualitative interview data to help to understand these results. During interviews, some students describe mentally switching from a numeric to a pictorial representation easily and using this skill to solve the numeric item:

\myquote{I just looked at the values and saw it -- like I kind of picture if they have that little bar with their error bars to see if they overlap.}

\noindent This student essentially converted the numeral data into pictorial data in their mind and then used that representation to reason about the comparisons. Using this skill of mentally changing representations, they were able to answer both items correctly.  This finding is similar to ones from Kohl et al. \cite{Kohl2005} and Weliweriya et al. \cite{Weliweriya2017}, in which students were often able to switch between different representations when forming mental models of data.

\vspace{-3mm}
\subsection{Individual answer analysis}
\vspace{-3mm}

In addition to comparing how well students scored on each question, we want to look at which answer options students choose to gain more insight into student reasoning. We determined how many students selected each of the seven answer options (due to the multiple response nature of the question, students could select multiple options, hence we do not expect these numbers to add up to 100\%). Table \ref{table:percentstuds} shows these data with 95\% confidence intervals. 

\begin{table*}[!ht]
    \caption{Percentage of students who selected each answer option with 95\% confidence interval}
    \label{table:percentstuds}
    \begin{tabularx}{\linewidth}{>{\hsize=1.4\hsize\linewidth=\hsize}X
>{\hsize=.6\hsize\linewidth=\hsize}X | >{\hsize=1.4\hsize\linewidth=\hsize}X >{\hsize=.6\hsize\linewidth=\hsize}X}
\hline\hline

Numeric  Representation (NRI) &  Percent of Students & Pictorial Representation (PRI) & Percent of Students \\ 
 & N = 670 & & N = 670 \\ \hline
\textbf{A. } $3.71 \pm 0.06 $ & 58 $\pm$ 4 & \textbf{A. } \includegraphics[width=.25\textwidth]{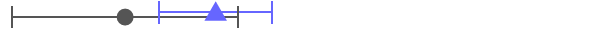} & 72 $\pm$ 3 \\ 
\textbf{B. } $3.71 \pm 0.17 $& 66 $\pm$ 4 & \textbf{B. } \includegraphics[width=.25\textwidth]{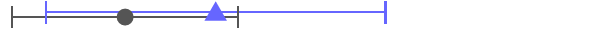} & 79 $\pm$ 3 \\ 
\textbf{C. } $3.76 \pm 0.06 $& 45 $\pm$ 4 & \textbf{C. } \includegraphics[width=.25\textwidth]{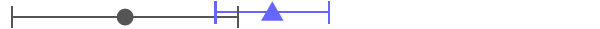} & 46 $\pm$ 4\\ 
\textbf{D. } $3.76 \pm 0.17 $& 55 $\pm$ 4  & \textbf{D. } \includegraphics[width=.25\textwidth]{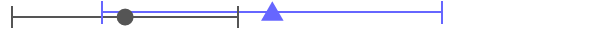} & 69 $\pm$ 4 \\ 
\textbf{E. } $3.91 \pm 0.06 $& 10 $\pm$ 2 & \textbf{E. } \includegraphics[width=.25\textwidth]{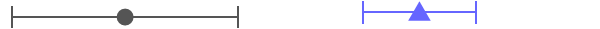} & 1.3 $\pm$ 0.9\\ 
\textbf{F. }\, $3.91 \pm 0.17 $& 15 $\pm$ 3 & \textbf{F. }\, \includegraphics[width=.25\textwidth]{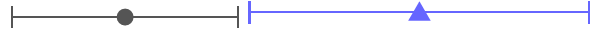} & 7 $\pm$ 2 \\ 
\textbf{G. }\, None of these agree with my data & 6 $\pm$ 2 & \textbf{G. }\, None of these agree with my data & 1.5 $\pm$ 0.9 \\
\hline \hline
    \end{tabularx}
\end{table*}

For both the PRI and the NRI, students most commonly select B, in which the means of both measurements lie within each other's error bars. The second most common choices were A and D, in which the error bars of only one of the measurements overlaps with the mean of the other measurement. This shows that, frequently, students require one of the means to be within the error bars of another measurement, as opposed to accepting error bar overlap as agreement between two measurements with uncertainty. 

Again from Table \ref{table:percentstuds}, many more students selected answer option E for the NRI than the PRI; this answer option is the only one where the two measurements definitely do not agree. Students identify this disagreement more frequently when presented with the data pictorially, where it is clear that the error bars are very far from one another, rather than when presented with this same data numerically. During interviews, one student selected all answer options (aside from ``None of the above'') on the NRI, and said:

\myquote{Honestly I would just say all of them... that's still at the end of the day what they got... We don't have enough data to say like `no yours are all wrong because they don't exactly match ours' because there are a lot of factors that could have altered their numbers and their uncertainty. I know that's a very idealized way of thinking about science.}

\noindent However, this student provided expert-like reasoning regarding overlap of the full range of each measurement when correctly answering the PRI, showing a clear difference in thinking about measurement comparison between the two representations.

\begin{figure}[!ht]
    \centering
    \includegraphics[width=.43\textwidth]{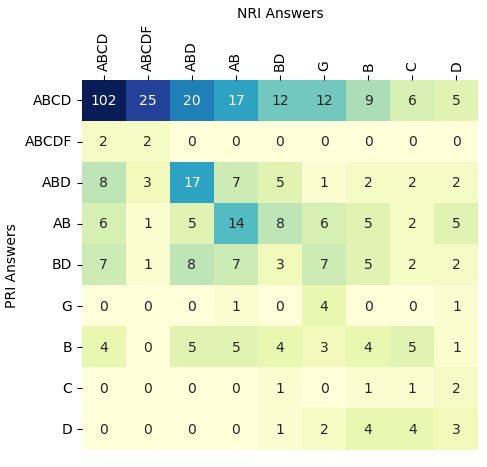}
    \caption{Heat map showing the most common answer combinations for the NRI and PRI [N = 409]. Answer combinations ABCD and ABCDF were marked as correct; no other combinations earned credit. Diagonal elements indicate students who answered identically to both the NRI and PRI, and off-diagonal elements indicate students who answered the items differently.}
    \label{fig:heatmap}
\end{figure}

Knowing the most commonly selected answer options allows us to delve further into common incorrect answer \textit{combinations} and reasonings for these choices. Figure \ref{fig:heatmap} shows a heat map of the most common answer combinations to each of the two questions (representing 409 of the 670 total student answers). The diagonal represents students who chose the same answer options for both the NRI and the PRI; the off-diagonal elements are students who selected different answers for each of these items.

One of the more common incorrect combinations on both items is `AB' [NRI: 54/670 = (8$\pm$ 2)\%, PRI: 79/670 = (12$\pm$ 2)\%]. This incorrect response aligns with students who consider their measurement more important in some sense, and therefore believe that the other groups' mean must be within their own error bars in order the measurements to agree with one another as compared to the other way around (requiring their mean to be within the other measurement's error bars), as would be indicated by the selection of `BD' [NRI: 45/670 = (7$\pm$ 2)\%, PRI: 63/670 = (9$\pm$ 2)\%].

For example, one student who selected only `AB' on the numeric item gave said:
 
 \myquote{For the other four groups, the uncertainties for their values did not put them in the same range as my values with its uncertainty so I don't believe they agree with my value.}
 
 \noindent In other words, when comparing numeric measurements with uncertainty, they placed more weight on their own measurement --- in order for agreement to occur, the uncertainty of the other measurement had to encompass their own mean. When solving this problem, they only added and subtracted their uncertainty to their own value and then selected the two answers whose means fell within that range; they ignored the uncertainties in the measurements in the answer options. However, we note that When answering the PRI, this same student selected a correct response of `ABCD', and provided expert-like reasoning. Thus, their reasoning changed with representation. 
 
 This theme of placing more importance on their own measurements frequently appeared in student interviews. Occasionally it was present when students provided reasoning for the PRI, but it was more typically found in student answers to the NRI.



 Another common incorrect answer for both items is ABD [NRI: 57/670 = (9 $\pm$ 2)\%; PRI: 60/670 = (10$\pm$2)\%]. In this line of incorrect reasoning, students did not consider answer option C to be correct, in which just the error bars overlap: they required at least one of the means to be within the error bars of the other measurement in order for agreement between measurements to occur.

Student reasoning from interviews supports this interpretation. For example, one student interviewed selected `ABD' on the PRI because:

\myquote{Not only do a large portion of their error bars overlap, it also contains the measurement itself,}

\noindent when referring to answer option B and D. They then chose A because:

\myquote{I would include [A] because now that measurement is included in mine, but [C] I am not sure about because... I don't necessarily know for sure they agree.}

\noindent In this example, the student did not consider answer option C to show agreement despite the error bar overlap - instead, they placed additional emphasis on requiring the mean to be included in at least one of the uncertainties of the other measurement.

Figure 2 also shows that very few students chose `ABCDF' for the PRI [4/670, or (0.60$\pm$ 0.06)\%], but many more students chose this for the NRI [the heat map shows 32 of the 35/670 = (5 $\pm$ 2)\% students who chose this option]. In the PRI, answer option F is one in which the error bars do not overlap, but are very close to each other, showing that agreement might be possible, hence why selection of F was not considered when scoring this item -- this option's correctness largely depends on which guidelines instructors teach students. Additionally, interview data showed mixed reasoning for students who selected this option.

\vspace{-4mm}
\section{Conclusions \& Takeaways\label{sec:conclusion}}
\vspace{-4mm}

Overall, students performed better on the PRI than the NRI, showing a more expert-like understanding of measurement comparison when presented with a pictorial format. However, students' did not perform as well as desired on either item, indicating room for improvement in teaching this important skill to students. Only about 40\% of students correctly identified whether measurements with uncertainties agree with one another in a pictorial format, and this drops to only about 25\% when presented numerically instead. Since many scientific papers generally provide numbers with uncertainties for measurements, this is a valuable skill needed in their future scientific careers to interpret experimental results. It is also vital for students to be able to work with many representations of data and convert between them. This study suggests that having students work with multiple representations, and convert between them, could be beneficial for developing expertise with measurement uncertainty and comparing measurements.

In future work, we will examine pre-post gains across this objective by examining scores prior to, and after, instruction in introductory laboratory courses. Additionally, we will explore other research directions using SPRUCE data, such as students' ideas around accuracy and precision and their ability to propagate errors to obtain an uncertainty in a calculated quantity. Finally, we will examine the alignment of student performance on SPRUCE with a variety of variables, including race, gender, institution type, and instructional methods.
\vspace{-7mm}
\acknowledgments{ \vspace{-5mm} This work is supported by NSF DUE 1914840, DUE 1913698, and PHY 1734006.. The authors wish to thank Marcos D. Caballero, Rachel Henderson, and Benjamin Pollard for their help in developing SPRUCE, as well as the students and instructors who participated in this study.}

\bibliographystyle{ieeetr}
\bibliography{bibfile} 

\begin{thebibliography}{10}

\bibitem{EP3}
``Guide to instructional laboratories and experimental skills.''
\newblock Accessed on January 18, 2023, Available at
  \url{https://ep3guide.org/guide-overview/instructional-laboratories-and-experimental-skills}.

\bibitem{Campbell2005}
B.~Campbell, F.~Lubben, A.~Buffler, and S.~Allie, ``Teaching scientific
  measurement at university: understanding students' ideas and laboratory
  curriculum reform,'' {\em Monograph, African Journal of Research in
  Mathematics, Science and Mathematics Education}, 2005.

\bibitem{Holmes2015}
N.~G. Holmes, C.~E. Wieman, and D.~A. Bonn, ``Teaching critical thinking,''
  {\em Proceedings of the National Academy of Sciences}, vol.~112,
  pp.~11199--11204, Sept. 2015.

\bibitem{Holmes2015_3}
N.~G. Holmes and D.~A. Bonn, ``{Quantitative Comparisons to Promote Inquiry in
  the Introductory Physics Lab},'' {\em The Physics Teacher}, vol.~53,
  pp.~352--355, 09 2015.

\bibitem{Stein2018}
M.~M. Stein, E.~M. Smith, and N.~G. Holmes, ``Confirming what we know:
  {Understanding} questionable research practices in intro physics labs,'' Dec.
  2018.

\bibitem{Quinn2018}
K.~N. Quinn, C.~E. Wieman, and N.~G. Holmes, ``Interview {Validation} of the
  {Physics} {Lab} {Inventory} of {Critical} thinking ({PLIC}),'' pp.~324--327,
  Mar. 2018.

\bibitem{Kok2022}
K.~W. Kok, {\em Certain about uncertainty}.
\newblock PhD thesis, Humboldt-Universit{\"a}t zu Berlin,
  Mathematisch-Naturwissenschaftliche Fakult{\"a}t, 2022.

\bibitem{Pollard2021}
B.~Pollard, R.~Hobbs, R.~Henderson, M.~D. Caballero, and H.~Lewandowski,
  ``Introductory physics lab instructors' perspectives on measurement
  uncertainty,'' {\em Physical Review Physics Education Research}, vol.~17,
  p.~010133, May 2021.

\bibitem{Vignal2023}
M.~Vignal, G.~Geschwind, B.~Pollard, R.~Henderson, M.~D. Caballero, and H.~J.
  Lewandowski, ``Survey of physics reasoning on uncertainty concepts in
  experiments: an assessment of measurement uncertainty for introductory
  physics labs,'' Feb. 2023.
\newblock arXiv:2302.07336 [physics].

\bibitem{Mislevy2005}
R.~Mislevy and M.~Riconscente, ``Evidence-{Centered} {Assessment} {Design}:
  {Layers}, {Structures}, and {Terminology},'' Jan. 2005.

\bibitem{Vignal2022}
M.~Vignal, K.~D. Rainey, B.~R. Wilcox, M.~D. Caballero, and H.~J. Lewandowski,
  ``Affordances of {Articulating} {Assessment} {Objectives} in {Research}-based
  {Assessment} {Development},'' pp.~475--480, Sept. 2022.

\bibitem{Kohl2005}
P.~B. Kohl and N.~D. Finkelstein, ``Student representational competence and
  self-assessment when solving physics problems,'' {\em Physical Review Special
  Topics - Physics Education Research}, vol.~1, p.~010104, Oct. 2005.

\bibitem{Kohl2006}
P.~B. Kohl and N.~D. Finkelstein, ``Effects of representation on students
  solving physics problems: {A} fine-grained characterization,'' {\em Physical
  Review Special Topics - Physics Education Research}, vol.~2, p.~010106, May
  2006.

\bibitem{Kohl2007}
P.~B. Kohl, D.~Rosengrant, and N.~D. Finkelstein, ``Strongly and weakly
  directed approaches to teaching multiple representation use in physics,''
  {\em Physical Review Special Topics - Physics Education Research}, vol.~3,
  p.~010108, June 2007.

\bibitem{Kohl2008}
P.~B. Kohl and N.~D. Finkelstein, ``{Patterns} of multiple representation use
  by experts and novices during physics problem solving,'' 2008.

\bibitem{Susac2017}
A.~Susac, A.~Bubic, P.~Martinjak, M.~Planinic, and M.~Palmovic, ``Graphical
  representations of data improve student understanding of measurement and
  uncertainty: An eye-tracking study,'' {\em Phys. Rev. Phys. Educ. Res.},
  vol.~13, p.~020125, Oct 2017.

\bibitem{Engelhardt2009}
P.~V. Engelhardt, ``An {Introduction} to {Classical} {Test} {Theory} as
  {Applied} to {Conceptual} {Multiple}-choice {Tests},'' Apr. 2009.

\bibitem{Ding2009}
L.~Ding and R.~Beichner, ``Approaches to data analysis of multiple-choice
  questions,'' {\em Physical Review Special Topics - Physics Education
  Research}, vol.~5, p.~020103, Sept. 2009.

\bibitem{Mann1947}
H.~B. Mann and D.~R. Whitney, ``On a {Test} of {Whether} one of {Two} {Random}
  {Variables} is {Stochastically} {Larger} than the {Other},'' {\em The Annals
  of Mathematical Statistics}, vol.~18, pp.~50--60, Mar. 1947.

\bibitem{Kirk2008}
R.~E. Kirk, {\em Statistics: {An} {Introduction}}, pp.~281--282.
\newblock Cengage Learning, 2008.

\bibitem{Cohen1988}
J.~Cohen, {\em Statistical {Power} {Analysis} for the {Behavioral} {Sciences}},
  pp.~20--27.
\newblock L. Erlbaum Associates, 1988.

\bibitem{Weliweriya2017}
N.~Weliweriya, T.~Huynh, and E.~Sayre, ``Standing fast: Translation among
  durable representations using evanescent representations in upper-division
  problem solving,'' pp.~432--435, Oct. 2017.

\end{thebibliography}

\end{document}